\documentclass[12pt]{article}
\usepackage{amsmath}
\usepackage{graphicx}
\usepackage{enumerate}
\usepackage{natbib}
\usepackage{multibib}
\usepackage{url} 

\usepackage{amsthm, amssymb, bm, bbm}
\usepackage{mathtools}
\usepackage{subcaption}
\usepackage{booktabs}
\usepackage{newfloat}
\usepackage{framed}
\usepackage[usenames,dvipsnames,svgnames,table]{xcolor}
\usepackage[pdfencoding=auto, psdextra,
            colorlinks=true,
            linkcolor=red,
            urlcolor=black,
            citecolor=MidnightBlue!70,
            hypertexnames=false]{hyperref}

\usepackage{setspace}

\makeatletter
\newcommand{\phantomlabel}[2]{
    \protected@write\@auxout{}{
        \string\newlabel{#2}{
            {\@currentlabel#1}{\thepage}
            {\@currentlabel#1}{#2}{}
        }
    }
    \hypertarget{#2}{}
}
\makeatother

\allowdisplaybreaks
\def\bal#1\eal{\begin{align}#1\end{align}}
\def\balnn#1\ealnn{\begin{align*}#1\end{align*}}

\DeclarePairedDelimiter\set{\{}{\}}
\DeclarePairedDelimiterX{\norm}[1]{\lVert}{\rVert}{#1}
\DeclarePairedDelimiterX{\abs}[1]{\lvert}{\rvert}{#1}
\DeclarePairedDelimiterX{\floor}[1]{\lfloor}{\rfloor}{#1}
\DeclarePairedDelimiterX{\ceil}[1]{\lceil}{\rceil}{#1}

\DeclarePairedDelimiterX{\expectarg}[1]{[}{]}{%
    \ifnum\currentgrouptype=16 \else\begingroup\fi
    \activatebar#1
    \ifnum\currentgrouptype=16 \else\endgroup\fi
}

\DeclarePairedDelimiterX{\variancearg}[1]{(}{)}{%
    \ifnum\currentgrouptype=16 \else\begingroup\fi
    \activatebar#1
    \ifnum\currentgrouptype=16 \else\endgroup\fi
}

\DeclarePairedDelimiterX{\klarg}[1]{(}{)}{%
    \ifnum\currentgrouptype=16 \else\begingroup\fi
    \activatedoublebar#1
    \ifnum\currentgrouptype=16 \else\endgroup\fi
}

\newcommand{\innermid}{\nonscript\;\delimsize\vert\nonscript\;}
\newcommand{\activatebar}{%
    \begingroup\lccode`\~=`\|
    \lowercase{\endgroup\let~}\innermid
    \mathcode`|=\string"8000
}

\newcommand{\doublemid}{\nonscript\;\delimsize\vert\delimsize\vert\nonscript\;}
\newcommand{\activatedoublebar}{%
    \begingroup\lccode`\~=`\|
    \lowercase{\endgroup\let~}\doublemid
    \mathcode`|=\string"8000
}

\newcommand{\iidsim}{\overset{\text{iid}}\sim}
\newcommand{\indsim}{\overset{\text{ind.}}\sim}
\newcommand{\Bern}{\operatorname{Bernoulli}}

\newcommand{\Reals}[1]{\mathbb{R}^{#1}}

\newcommand{\bY}{\mathbf{Y}}
\newcommand{\by}{\mathbf{y}}

\newcommand{\bz}{\mathbf{z}}

\newcommand{\bO}{\mathbf{O}}

\newcommand{\bnu}{\boldsymbol\nu}

\newcommand{\bmu}{\boldsymbol\mu}

\newcommand{\btheta}{\boldsymbol\theta}

\begin{document}

\def\spacingset#1{\renewcommand{\baselinestretch}%
{#1}\small\normalsize} \spacingset{1}

\title{\bf \Large A Latent Space Approach to Inferring Distance-Dependent Reciprocity in Directed Networks}

\author{
Joshua Daniel Loyal, Xiangyu Wu, Jonathan R. Stewart \\[0.5em] 
Department of Statistics, Florida State University
}

\date{}
\maketitle

\begin{abstract}
Reciprocity, or the stochastic tendency for actors to form mutual relationships, is an essential characteristic of directed network data. Existing latent space approaches to modeling directed networks are severely limited by the assumption that reciprocity is homogeneous across the network. In this work, we introduce a new latent space model for directed networks that can model heterogeneous reciprocity patterns that arise from the actors' latent distances. Furthermore, existing edge-independent latent space models are nested within the proposed model class, which allows for meaningful model comparisons. We introduce a Bayesian inference procedure to infer the model parameters using Hamiltonian Monte Carlo. Lastly, we use the proposed method to infer different reciprocity patterns in an advice network among lawyers, an information-sharing network between employees at a manufacturing company, and a friendship network between high school students.
\end{abstract}


\onehalfspacing

\section{Introduction}\label{sec:intro}

Networks, which consist of nodes (i.e., entities) and edges (i.e., pairwise relationships), are frequently analyzed to quantify the stochastic tendency of specific relationships to form in a population. These relationships are often directed, meaning the relationship between two nodes can be asymmetric. For example, in a friendship network, node $i$ may consider node $j$ a friend even if the reverse does not hold. As such, a stochastic tendency at the core of many scientific problems involving directed networks is {\it reciprocity} or the tendency of nodes to form mutual (or symmetric) relationships. For instance, reciprocity influences how fast infectious diseases spread through human contact networks~\citep{li2013} or information propagates on online social networks~\citep{zhu2014}. Furthermore, real-world networks demonstrate distinct reciprocity patterns that can elucidate the mechanisms behind network formation~\citep{garlaschelli2004}. In particular, \citet{jiang2015} found that online social networks tend to have higher levels of reciprocity than biological or communication networks. As such, flexible statistical models of reciprocity are essential for rigorous scientific understanding.

Various statistical models have been developed to quantify reciprocity in directed network data. Influential models are the $p_1$ model~\citep{holland1981}, the $p_2$ model~\citep{duijn2004}, certain exponential random graph models (ERGMs)~\citep{frank1986, lusher2013}, directed stochastic block models~\citep{holland1983, wang1987}, and the additive and multiplicative effects network (AMEN) model~\citep{hoff2021}. These models include a single parameter for reciprocity that is homogeneous across the network, that is, they assume the tendency to reciprocate is the same for all relationships. As such, these models fail to account for heterogeneous reciprocity patterns that vary at the edge level. To account for reciprocity levels that vary according to network subpopulations, \citet{vu2013} proposed a mixture of ERGMs. In addition, \citet{zijlstra2006} introduced an extension to the $p_2$ model that expressed reciprocity as a function of exogenous covariates. However, no existing method accounts for edge-level heterogeneous reciprocity patterns explained by endogenous network properties. 

This work is concerned with modeling heterogeneous reciprocity patterns using latent space models (LSMs) for networks~\citep{hoff2002}, which have been widely adopted due to their flexibility and interpretability. LSMs represent each node $i$ in the network with a latent position $\bz_i \in \Reals{d}$ in a low-dimensional Euclidean space with the assumption that nodes that are closer together in latent space are more likely to form a connection in the network. The key idea behind LSMs is that the distances between nodes in the lower-dimensional latent space are responsible for the stochastic tendencies toward certain relationships in the network. Since their introduction, various LSMs have been developed to capture important stochastic tendencies~\citep{handcock2007, krivitsky2009, hoff2005, ma2020}, such as homophily, degree heterogeneity, clustering, and transitivity (i.e., triad formation). Despite this growing body of work, LSMs rarely directly account for reciprocity. In fact, to the best of our knowledge, the only existing LSM with reciprocity is the AMEN model, which includes a single parameter to account for homogeneous reciprocity.

The contributions of this article are two-fold: (1) we introduce the concept of {\it distance-dependent reciprocity} to parsimoniously describe heterogeneous reciprocity patterns in directed network data, and (2) we develop an LSM to rigorously model and infer this new reciprocity type. Specifically, we define distance-dependent reciprocity as the stochastic tendency for two nodes to reciprocate a relationship based on their distance in latent space. Intuitively, distance-dependent reciprocity asserts that a node may reciprocate edges more (or less) with other similar nodes, e.g., mutual friendships may be more (or less) common for nodes of the same gender. Building off the $p_1$ model, we introduce an LSM with distance-dependent reciprocity that maintains LSMs' existing advantages while also flexibly modeling this new heterogeneous reciprocity pattern. Furthermore, we develop a Bayesian inference procedure to determine the type of distance-dependent reciprocity present in observed network data. Overall, the proposed model provides researchers with deeper insights into the structure of directed networks than existing LSMs.

\begin{figure*}[h]
\centerline{\includegraphics[height=0.3\textheight, width=\textwidth, keepaspectratio]{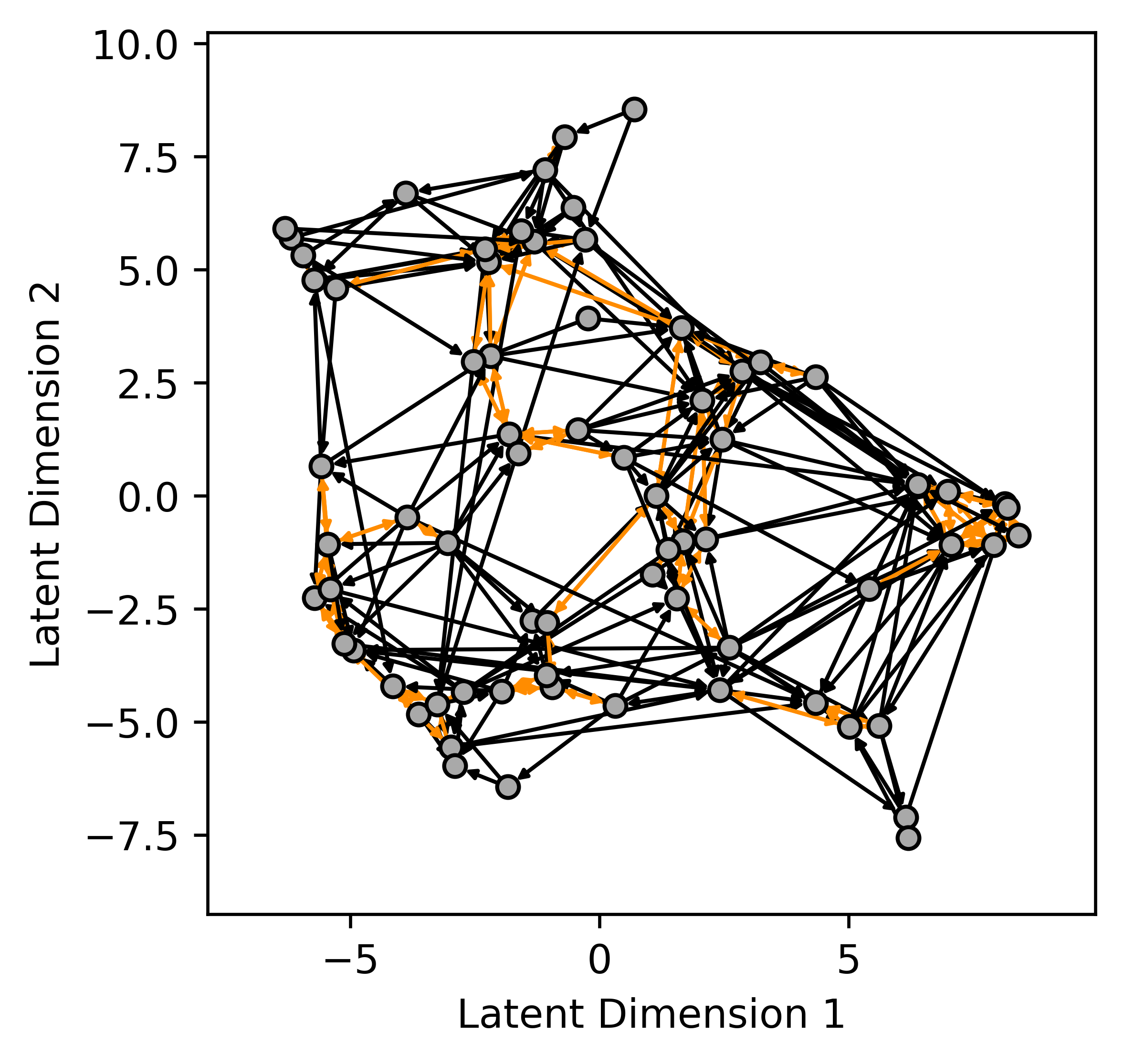}}
\caption{Network of friendship relations between high school students visualized using the LSM of \citet{krivitsky2009} with $d = 2$ latent dimensions. Reciprocated edges (orange) occur more often at small distances than unreciprocated edges (black).}
\label{fig:illinois_net}
\end{figure*}

For further motivation, we provide an example of distance-dependent reciprocity in Figure~\ref{fig:illinois_net}, which displays a network of friendship relations between high school students collected by \citet{coleman1964}. The network is visualized using the latent space model of \citet{krivitsky2009} with $d = 2$ latent dimensions. For a complete description of this model, see Equation~(\ref{eq:ind_lsm}) in Section~\ref{sec:existing}. It is clear from this network embedding that mutual (orange) ties occur more often over smaller distances in latent space compared to unreciprocated (black) ties. Previous LSMs do not account for this interaction between a pair of nodes' latent similarity and their tendency to reciprocate. As such, existing models fail to capture this important network characteristic.

The remainder of this article is structured as follows. Section~\ref{sec:model} introduces the proposed LSM with distance-dependent reciprocity, which generalizes a few existing methods. Section~\ref{sec:estimation} proposes a Bayesian approach to estimating the model parameters using Markov chain Monte Carlo (MCMC). Section~\ref{sec:simulation} demonstrates the ability of the proposed MCMC algorithm to recover the true model parameters on simulated networks. Then, we use the methodology to infer different types of distance-dependent reciprocity in three real-world networks from the social sciences in Section~\ref{sec:applications}. Lastly, Section~\ref{sec:discussion} contains a discussion.

\section{An LSM with Distance-Dependent Reciprocity}\label{sec:model}

\subsection{Directed Network Data}

Consider a binary directed network on $n$ nodes. We represent the network as a binary $n \times n$ adjacency matrix $\bY$ with random entries $Y_{ij} \in \set{0, 1}$ for $1 \leq i,j \leq n$, such that $Y_{ij} = 1$ if there exists a directed relation from node $i$ to node $j$ and is zero otherwise. We assume there are no self-loops, so the diagonal elements $Y_{ii} = 0$.  Each node pair $(i, j)$ is associated with a pair of edge variables $D_{ij} = (Y_{ij}, Y_{ji})$ known as a dyad. A dyad can take on four values $D_{ij} \in \set{(1,1), (1,0), (0,1), (0,0)} = \{``i \leftrightarrow j",$ $``i \rightarrow j", ``j \rightarrow i", ``i - j"\}$, which indicate the four types of directed relations between nodes $i$ and $j$. In particular, ``$i \leftrightarrow j$" denotes a mutual or reciprocated relationship, ``$i \rightarrow j"$ and ``$j \rightarrow i"$ denote asymmetric or unreciprocated relationships, and ``$i - j"$ denotes no relationship. In what follows, we let $\by \in \set{0, 1}^{n \times n}$ denote the observed adjacency matrix with $(i,j)$-th entry $y_{ij}$.

\subsection{Background: A Dyad-Independent Exponential Family}

Our proposed latent space model is a generalization of the $p_1$ model of \citet{holland1981} for modeling reciprocity and degree heterogeneity in directed networks. As in the $p_1$ model, we assume that the dyads $\set{D_{ij} : 1\leq i<j \leq n}$ are conditionally independent given a set of model parameters $\btheta$, so that
\begin{align}
    \mathbb{P}_{\btheta}(\bY = \by) &= \prod_{i < j} \mathbb{P}_{\btheta}(Y_{ij} = y_{ij}, Y_{ji} = y_{ji}) \nonumber \\ 
                          &= \prod_{i < j} p_{i\leftrightarrow j}^{y_{ij} y_{ji}} \,   p_{i\rightarrow j}^{y_{ij} (1 - y_{ji})}  \, p_{j \rightarrow i}^{(1 - y_{ij}) y_{ji}} \, p_{i - j}^{(1 - y_{ij})(1 - y_{ji})}, \label{eq:lik_multinom}
\end{align}
where  $p_{i \leftrightarrow j} = \mathbb{P}_{\btheta}(Y_{ij} = 1, Y_{ji} = 1)$, $p_{i \rightarrow j} = \mathbb{P}_{\btheta}(Y_{ij} = 1, Y_{ji} = 0)$, $p_{j \rightarrow i} = \mathbb{P}_{\btheta}(Y_{ij} = 0, Y_{ji} = 1)$, and $p_{i-j} = \mathbb{P}_{\btheta}(Y_{ij} = 0, Y_{ji} = 0)$ denote the probabilities of observing each dyad value so that $p_{i \leftrightarrow j} + p_{i  \rightarrow j} + p_{j \rightarrow i} + p_{i - j} = 1$. Equation~(\ref{eq:lik_multinom}) is often expressed in terms of the natural parameters of an exponential family:
\begin{equation}\label{eq:lik_expfam}
    \mathbb{P}_{\btheta}(\bY = \by) \propto \prod_{i < j} \exp(\mu_{ij} y_{ij} + \mu_{ji} y_{ji} + \rho_{ij} y_{ij} y_{ji}), 
\end{equation}
where
\[
    \mu_{ij} = \log(p_{i\rightarrow j}\, / \, p_{i-j}) = \text{logit}\{\mathbb{P}_{\btheta}(Y_{ij} = 1 \mid Y_{ji} = 0)\},
\]
and
\[
    \rho_{ij} = \log\left(\frac{p_{i \leftrightarrow j} \ p_{i-j}}{p_{i \rightarrow j} \ p_{j \rightarrow i}}\right) = \log \frac{\text{odds}_{\btheta}(Y_{ij} = 1 \mid Y_{ji} = 1)}{\text{odds}_{\btheta}(Y_{ij} = 1 \mid  Y_{ji} = 0)}.
\]
In the previous expressions, $\mu_{ij}$ is the log odds of observing an unreciprocated dyad given that $Y_{ji} = 0$. Furthermore, $\rho_{ij}$ is a log odds ratio that quantifies the degree of reciprocity between nodes $i$ and $j$ in the sense that if $\rho_{ij}$ is positive, then the odds of observing $Y_{ij} = 1$ given $Y_{ji} = 1$ (i.e., a mutual relationship) is higher than the odds of observing $Y_{ij} = 1$ given $Y_{ji} = 0$ (i.e., an asymmetric relationship). That is $\rho_{ij} > 0$ indicates that node $i$ is more likely to reciprocate an existing relationship than enter into an unreciprocated relationship with node $j$. Note that $\rho_{ij}$ is symmetric so that $\rho_{ij} = \rho_{ji}$. Furthermore, $\rho_{ij} = 0$ indicates that the edge variables $Y_{ij}$ and $Y_{ji}$ are independent. As defined, the model is over-parameterized, so restrictions must be placed on the natural parameters. In what follows, we propose a set of restrictions designed to model degree heterogeneity, transitivity, and, most importantly, distance-dependent reciprocity in directed network data.

\subsection{The Proposed Latent Space Model}
Now, we describe our latent space model for inferring distance-dependent reciprocity, which builds on Equation~(\ref{eq:lik_expfam}).  We adopt the latent space modeling approach~\citep{hoff2002} and assign each node $i$ a latent position $\bz_i \in \Reals{d}$ in a low-dimensional $(d \ll n)$ Euclidean space. As such, we parameterize our model in terms of distances in latent space as follows:
\begin{equation}\label{eq:rlsm}
    \mu_{ij} = s_i + r_j - \norm{\bz_i - \bz_j}_2, \qquad  \rho_{ij} = \rho + \phi \norm{\bz_i - \bz_j}_2,
\end{equation}
which we interpret in terms of the following equations
\begin{align}
\text{logit}\{\mathbb{P}_{\btheta}(Y_{ij} = 1 \mid Y_{ji} = y_{ji})\} &= \begin{cases}
    s_i + r_j - \norm{\bz_i - \bz_j}_2 & y_{ji} = 0 \\
    s_i + r_j + \rho + (\phi - 1) \norm{\bz_i - \bz_j}_2 & y_{ji} = 1, 
\end{cases} \label{eq:rlsm_logit} \\
    \intertext{and}
    \log \frac{\text{odds}_{\btheta}(Y_{ij} = 1 \mid Y_{ji} = 1)}{\text{odds}_{\btheta}(Y_{ij} = 1 \mid Y_{ji} = 0)} &= \rho + \phi \norm{\bz_i - \bz_j}_2. \label{eq:dist_rep}
\end{align}
In the expression for the conditional log odds $\mu_{ij}$, we assigned each node a sender parameter $s_i \in \Reals{}$ and a receiver parameter $r_i \in \Reals{}$ to capture degree heterogeneity in the network. Furthermore, the log odds ratio $\rho_{ij}$ is parameterized using a baseline reciprocity parameter $\rho \in \Reals{}$ and a distance coefficient $\phi \in \Reals{}$.  The proposed model has $n(d+2) + 2$ parameters $\btheta = \{\bz_1, \dots, \bz_n, s_1, \dots, s_n, r_1, \dots, r_n, \rho, \phi\}$,  which is much less than the $3n(n-1)/2$ parameters in Equation (\ref{eq:lik_expfam}). Note that when the latent distances are removed from Equation~(\ref{eq:rlsm}), the model reduces to the $p_1$ model of \citet{holland1981}.

Crucially, the proposed model exhibits distance-dependent reciprocity, meaning that the reciprocity between two nodes is a function of their distance in latent space. In particular, Equation~(\ref{eq:dist_rep}) demonstrates that the degree of reciprocity between two nodes is a linear function of $\norm{\bz_i - \bz_j}_2$. The baseline reciprocity $\rho$ is the degree of reciprocity between two nodes with the same latent positions. The distance coefficient $\phi$ quantifies how the degree of reciprocity between two nodes deviates from this baseline as their distance increases. Specifically, the tendency towards mutual ties decreases with distance for $\phi < 0$ and increases for $\phi > 0$. For $\phi = 0$, the degree of reciprocity is independent of distance and homogeneous across the network. 

We can also interpret the model parameters through the conditional log odds in Equation~(\ref{eq:rlsm_logit}). In both conditional log odds, the sender and receiver parameters appear as additive terms that capture out-degree and in-degree heterogeneity, respectively. The distant-dependent reciprocity $\rho_{ij} = \rho + \phi \norm{\bz_i - \bz_j}_2$ acts as an offset between the conditional log odds of an unreciprocated tie given $Y_{ji} = 0$ and a mutual tie given $Y_{ji} = 1$. In other words, adding $\rho_{ij}$ to the first line of (\ref{eq:rlsm_logit}) produces the second line. Consequently, the latent distance can affect the formation of an edge from node $i$ to node $j$ differently depending on whether there is a preexisting relationship from $j$ to $i$. When there is no preexisting relationship from $j$ to $i$, the log odds of an unreciprocated edge behaves like a standard latent space model, that is, it decreases directly with the distance $\norm{\bz_i - \bz_j}_2$. However, given an existing relationship from $j$ to $i$, the latent distance appears in the log odds of a mutual edge multiplied by an additional factor of $\phi - 1$. 

Based on the previous observations, we can also interpret $\phi$ in terms of its relative effect on the two conditional log odds in (\ref{eq:rlsm_logit}). When $\phi < 0$, the conditional log odds of mutual ties decrease more rapidly with distance than the conditional log odds of unreciprocated ties. This relative behavior results in smaller reciprocity levels at larger distances. When $\phi = 0$, the distances have the same effect on both conditional log odds, which results in homogeneous reciprocity throughout the network. For $0 < \phi  < 1$, the conditional log odds of mutual ties decays less rapidly with distance than the conditional log odds of unreciprocated ties, which results in higher reciprocity levels at larger distances. When $\phi = 1$, the conditional log odds of a mutual edge does not depend on the distances. This behavior describes networks where once an edge exists between two actors, the propensity to reciprocate that relationship does not depend on the actors' latent similarity. When $\phi > 1$, the conditional log odds of mutual ties increase with distance, opposite of their effect on unreciprocated ties, resulting in much higher reciprocity at longer distances.

\subsection{Connections to Existing Models}\label{sec:existing}

An important property of the proposed LSM with distance-dependent reciprocity is that it contains existing LSMs as special cases. A natural model for comparison is the edge-independent LSM of \citet{krivitsky2009}, which assumes that
\begin{align}
    \mathbb{P}_{\btheta}(\bY = \by) &= \prod_{i \neq j}\mathbb{P}_{\btheta}(Y_{ij} = y_{ij}), \quad \text{with}  \nonumber \\
    \text{logit}\{\mathbb{P}_{\btheta}(Y_{ij} = 1)\} &= s_i + r_j - \norm{\bz_i - \bz_j}_2. \label{eq:ind_lsm}
\end{align}
This edge-independent LSM is nested within the proposed model and is obtained when $\rho = \phi = 0$ so that $\rho_{ij} = 0$ for all dyads. In the case that $\rho \neq 0$ and $\phi = 0$, we obtain a dyad-independent LSM with homogeneous reciprocity captured by a bivariate Bernoulli model. This contrasts with the AMEN model~\citep{hoff2021}, which models homogeneous reciprocity using a bivariate probit model. Furthermore, when $\phi \neq 0$, we obtain a new class of models for heterogeneous reciprocity in directed network data.

Note that edge-independent LSMs can capture certain empirical heterogeneous reciprocity patterns, so a reasonable question is whether more complicated forms of reciprocity (e.g., homogeneous or distance-dependent) are needed to describe an observed network. The reason edge-independent LSMs can capture heterogeneous reciprocity is that the inhomogeneous edge-probabilities in Equation~(\ref{eq:ind_lsm}) induce networks that are locally dense and globally sparse in the sense that the number of connections will (on average) be higher in local regions of latent space relative to the density of the entire network. These localized densities allow LSMs to induce a higher stochastic tendency towards transitivity in networks since increasing the probability of edges in a local area also increases the probability of triangle formation. This property of LSMs has a similar effect on edge reciprocation. In this case, mutual edges are more probable at short distances in the latent space relative to long distances. However, this characteristic may be antithetical to certain networks where reciprocation may still be likely even when edge formation is unlikely, e.g., when mutual edges are more likely to form between nodes of different groups. For a real-world example of this phenomenon, see the information-sharing network in Section~\ref{sec:infoshare}. For this reason, we proposed the more general model in Equations (\ref{eq:lik_expfam})--(\ref{eq:dist_rep}).

This possible asymmetry between the propensity of edge formation and reciprocation has been observed and discussed with respect to sparse networks~\citep{krivitsky2015, stewart2019}. An example is the sparse Bernoulli model with reciprocation of \citet{krivitsky2015}, where the conditional log odds of edges are  
\begin{align}
    \mu_{ij} = \alpha - \log n, \quad \rho_{ij} &= \rho + \log n, \\
\label{eq:KrKo}
    \text{logit}\{\mathbb{P}_{\alpha, \rho}(Y_{ij} = 1 \mid Y_{ji} = y_{ji})\} &=
\begin{cases}
\alpha - \log n & y_{ji} = 0 \\
\alpha + \rho & y_{ji} = 1,
\end{cases}
\end{align}
with $\alpha, \rho \in \Reals{}$. Under this model, the marginal probability of an edge decreases with the size of the network $n$, resulting from the $-\log n$ term in the first line of \eqref{eq:KrKo}. In contrast, the conditional log odds of an edge is constant provided the edge is reciprocated, which is accomplished by adding an offset of $\rho+\log n$ to the first line of \eqref{eq:KrKo}, producing the second line. The analog for dyad-independent LSMs is a distance-dependent reciprocity term, which produces a similar offset for reciprocated edges based on the distances between two nodes in latent space. Specifically, in the proposed model under Equation~(\ref{eq:rlsm_logit}), this offset is $\rho + \phi \, \norm{\bz_i - \bz_j}_2$.

The benefit of the proposed model is that inference on its parameters can easily distinguish between certain types of reciprocity patterns not explained by an edge-independent LSM. For example, the probability of a reciprocated edge may require an additional form of dependence, which occurs when $\rho \neq 0$ or $\phi \neq 0$. Moreover, as discussed above, reciprocation may still be likely even when edge formation is unlikely. Such a phenomenon is captured when $0 < \phi < 1$, so that $\rho_{ij}$ increases with distance, while the conditional log odds of an edge decreases with distance. Table~\ref{tab:model_types} summarizes various scenarios of interest. In the next section, we develop a Bayesian inference scheme to infer the type of distance-dependent reciprocity in an observed network.

\begin{center}
\begin{table*}[hb]
\centering\small
    \begin{tabular*}{\textwidth}{@{\extracolsep\fill}lll@{\extracolsep\fill}} \toprule
    \textbf{Parameters} & \textbf{Dependence Type} & \textbf{Reciprocity Type} \\
\midrule
    $\rho = 0, \phi = 0$ & Edge-Independent \citep{krivitsky2009} & None \\[0.25em]
    $\rho \neq 0, \phi = 0$ & Dyad-Independent & Homogeneous \\[0.25em]
    $\phi \neq 0$ &  Dyad-Independent & Distance-Dependent \\[0.25em]
    \midrule
    $\qquad\quad \phi < 0$ &  & Decreases with latent distance \\[0.25em]
    $\qquad\quad \phi > 0$ &  & Increases with latent distance \\
\bottomrule
\end{tabular*}
\caption{Types of dependence and reciprocity nested within the proposed model.}
\label{tab:model_types}
\end{table*}
\end{center}

\section{Bayesian Estimation}\label{sec:estimation}

\subsection{Prior Specification}\label{sec:priors}

In what follows, we take a Bayesian approach to estimation. To do so, we assign appropriate priors to the model parameters and develop a Markov chain Monte Carlo (MCMC) algorithm to sample from the parameter's posterior distribution. Based on the posterior, we can construct parameter estimates and infer the type of reciprocity in a given network.

To complete the Bayesian model, we assign priors to the model parameters. For the sender and receiver parameters, we place the following independent multivariate normal priors
\begin{equation}\label{eq:sr_prior}
    \begin{pmatrix}s_i \\ r_i\end{pmatrix} \iidsim N\left\{\mu_{sr} \mathbf{1}_2, \begin{pmatrix} \sigma_s^2 & \gamma_{sr} \sigma_s \sigma_r \\ \gamma_{sr} \sigma_s \sigma_r  & \sigma_r^2 \end{pmatrix}\right\} \qquad (i = 1, \dots, n),
\end{equation}
with mean $\mu_{sr} \in \Reals{}$, sender and receiver variances $\sigma^2_s, \sigma^2_r > 0$, and sender-receiver correlation $\gamma_{sr}  \in [-1,1]$. In the previous expression, we use $\mathbf{1}_2$ to denote a two-dimensional vector of ones and $\iidsim$ to denote independent and identically distributed. This prior is similar to the random effects structure imposed on the sender and receiver parameters in the $p_2$ model~\citep{duijn2004} and the AMEN model. \citet{krivitsky2009} used a simplified prior with zero correlation $\gamma_{sr} = 0$. A minor change is our inclusion of a non-zero mean $\mu_{sr}$. Unlike the previous methods, for identifiability, our parametrization of $\mu_{ij}$ in Equation~(\ref{eq:rlsm}) does not include an intercept, which would take the form $\mu_{ij} = \mu_{0} + s_i + r_j - \norm{\bz_i - \bz_j}_2$. As such, $\mu_{sr}$ adjusts the sender and receiver parameters' center to account for the overall density of the network. We place the following standard hyperpriors on this prior's parameters:
\begin{align}
    \mu_{sr} \sim N(0, \sigma^2_{\mu}), \quad \sigma_s^2 \sim \Gamma^{-1}(a_s, b_s), \quad \sigma_r^2 \sim \Gamma^{-1}(a_r,b_r), \quad \gamma_{sr} \sim \text{Uniform}[-1, 1],
\end{align}
where $\Gamma^{-1}(a,b)$ denotes an inverse-gamma distribution with shape $a$ and scale $b$, and $\text{Uniform}[-1,1]$ is the uniform distribution on $[-1,1]$. We set $a_s = b_s = a_r = b_r = 3/2$ and $\sigma_{\mu} = 10$ to induce relatively broad hyperpriors. 

Standard in the Bayesian LSM literature, we give the latent positions independent mean-zero multivariate normal priors:
\begin{equation}
\bz_i \iidsim N(\mathbf{0}_d, \sigma^2_z \mathbf{I}_d) \qquad (i = 1, \dots, n),
\end{equation}
where $\mathbf{0}_d$ and $\mathbf{I}_d$ denote a $d$-dimensional vector of zeros and the $d$-dimensional identity matrix, respectively. This prior is used in the original model of \citet{hoff2002} and the edge-independent model of \citet{krivitsky2009}. Similar to the sender and receiver variances, we assign $\sigma_z^2 \sim \Gamma^{-1}(a_z, b_z)$ with $a_z = b_z = 3/2$, which induces a broad prior. 

Lastly, we assign normal priors to the distant-dependent reciprocity parameters:
\begin{equation}\label{eq:dd_priors}
\rho \sim N(0, \sigma_{\rho}^2), \qquad \phi \sim N(0, \sigma_{\phi}^2).
\end{equation}
We set $\sigma_{\rho} = \sigma_{\phi} = 10$, so the priors are broad. Importantly, these priors are centered about zero, for which our model reduces to an edge-independent LSM.

\subsection{Posterior Inference}\label{sec:hmc}

We make inferences about the model parameters based on the joint posterior of $\btheta = \{\bz_1, \dots, \bz_n, s_1, \dots, s_n, r_1, \dots, r_n, \rho, \phi\}$ and $\bnu = \set{\mu_{sr}, \sigma_s^2, \sigma_r^2, \gamma_{sr}, \sigma_z^2}$, that is,
\[
    p(\btheta, \bnu \mid \bY = \by) \propto \mathbb{P}_{\btheta}(\bY = \by) \ p(\btheta \mid \bnu) \ p(\bnu),
\]
where the likelihood is given by the model in Equations (\ref{eq:lik_expfam})--(\ref{eq:dist_rep}) and $p(\btheta \mid \bnu)$ and $p(\bnu)$ are the hierarchical priors defined in Equations (\ref{eq:sr_prior})--(\ref{eq:dd_priors}). We sample from this posterior using Hamiltonian Monte Carlo (HMC) with adaptive tuning parameter selection~\citep{neal2011, hoffman2014} as implemented in NumPyro~\citep{phan2019}. Like many latent space models, the posterior is invariant to translations and rotations of the latent positions. We use the approach of Procrustes matching to resolve this non-identifiability. Specifically, we post-process the MCMC output by rotating and translating the latent positions to match a reference layout. For a detailed description of this procedure, see the original work by \citet{hoff2002}. We use the maximum a posteriori (MAP) estimator as the reference layout.

Lastly, since the model has a large number of parameters, a good initialization near the posterior mode can greatly reduce the number of iterations needed for the MCMC algorithm to converge. We use the following procedure to initialize the parameters:
\begin{enumerate}
    \item Multidimensional scaling is performed on the geodesic distances between the nodes according to the undirected network $\tilde{\by}$ with entries $\tilde{y}_{ij} = 1\set{y_{ij} + y_{ji} > 0}$ to get the initial latent positions $\bz_1^{(0)}, \dots, \bz_n^{(0)}$. These positions are then centered at the origin.
    \item The sender and receiver effects are initialized based on the observed out-degrees and in-degrees using a Bayesian estimator of proportions. Specifically, for $i = 1, \dots, n$, we set the initial sender and receive effects of node $i$ to the log odds of the posterior means of the probabilities under the following model:
        \begin{align*}
            &Y_{ij} \iidsim \Bern(p_i^{\text{out}}), \quad p_i^{\text{out}} \sim \text{Uniform}(0, 1) \quad (1 \leq  j \neq i \leq n), \\
            &Y_{ji} \iidsim \Bern(p_i^{\text{in}}), \quad p_i^{\text{in}} \sim \text{Uniform}(0, 1) \quad (1 \leq j \neq i \leq n),
        \end{align*}
        that is, we set $s_i^{(0)} = \text{logit}\set{(\sum_{j=1}^n y_{ij} + 1)/(n + 2)}$ and $r_i^{(0)} = \text{logit}\set{(\sum_{j=1}^n y_{ji} + 1) / (n + 2)}$ for $i = 1, \dots, n$. This approach was used to regularize the estimates for nodes with an in-degree or out-degree of zero. We set $\mu_{sr}^{(0)} = (1/2n) \sum_{i=1}^n (s_i^{(0)}+ r_i^{(0)})$ and then centered the initial sender and receiver effects about zero.
    \item The reciprocity parameters $\rho$ and $\phi$ were initialized be setting them to their estimates under the following logistic regression model 
        \begin{align*}
            &Y_{ij} \mid Y_{ji} = y_{ji} \indsim \Bern\{\mathbb{P}(Y_{ij} \mid Y_{ji} = y_{ji})\} \quad (1 \leq i < j \leq n), \\
            &\text{logit}\set{\mathbb{P}(Y_{ij} = 1 \mid Y_{ji} = y_{ji})} = \beta_0 + \rho y_{ji} + \phi \norm{\bz_i^{(0)} - \bz_j^{(0)}}_2 y_{ji},  
        \end{align*}
        where $\indsim$ denotes independently distributed and the latent positions are the initial estimates from Step 1. The above model implies the log odds ratio in Equation~(\ref{eq:dist_rep}).
        
    \item The remaining parameters, that is, $\sigma_s^2, \sigma_r^2, \gamma_{sr}, \sigma_z^2$, are initialized using Numpyro's default initialization scheme.
\end{enumerate}

\section{Simulation Study}\label{sec:simulation}
Here, we present a simulation study to assess the ability of the MCMC algorithm to recover the model parameters as the number of nodes increased. We generated 50 synthetic networks from the model in Equations (\ref{eq:lik_expfam})--(\ref{eq:rlsm}) with latent space dimension $d = 2$ and the number of nodes $n \in \{50, 100, 150, 200, 250\}$. The sender and receiver parameters of each node $s_i$ and $r_i$ were sampled independently from the multivariate normal prior in Equation~(\ref{eq:sr_prior}) with $\sigma^{2}_{s} = \sigma^{2}_{r} = 1$ and $\gamma_{sr} = 0.5$. Furthermore, we set $\mu_{sr}$ to fix the expected density of the observed networks to 0.2. We sampled the latent position of each node independently from a three-component Gaussian mixture, that is,  $\bz_i \iidsim (1/3) N(\bmu_1, \sigma^2\mathbf{I}_{2}) + (1/3) N(\bmu_2, \sigma^2\mathbf{I}_{2}) + (1/3) N(\bmu_3, \sigma^2\mathbf{I}_{2})$, with $\bmu_1 = (-1, 0)^{\top}$, $\bmu_2 = (1, 0)^{\top}$, $\bmu_3 = (0, 1)^{\top}$, and $\sigma^2 = 0.1$. For each network, we sampled the distance coefficient $\phi$ from a $\text{Uniform}[-1, 1]$.  Lastly, we set the baseline reciprocity parameter $\rho$ so that the average odds ratio, that is, $\rho_{ij}$ in Equation (\ref{eq:dist_rep}), equaled two. In all simulations, we estimated the posterior distributions using the MCMC algorithm outlined in Section \ref{sec:hmc} with $d = 2$ and hyperparameters set as described in Section \ref{sec:priors}. We ran the MCMC algorithm for 5,000 iterations after a burn-in of 2,500 iterations. We estimated all model parameters using their posterior mean.

\begin{figure}[htbp]
    \centering
    \includegraphics[width=\textwidth, keepaspectratio]{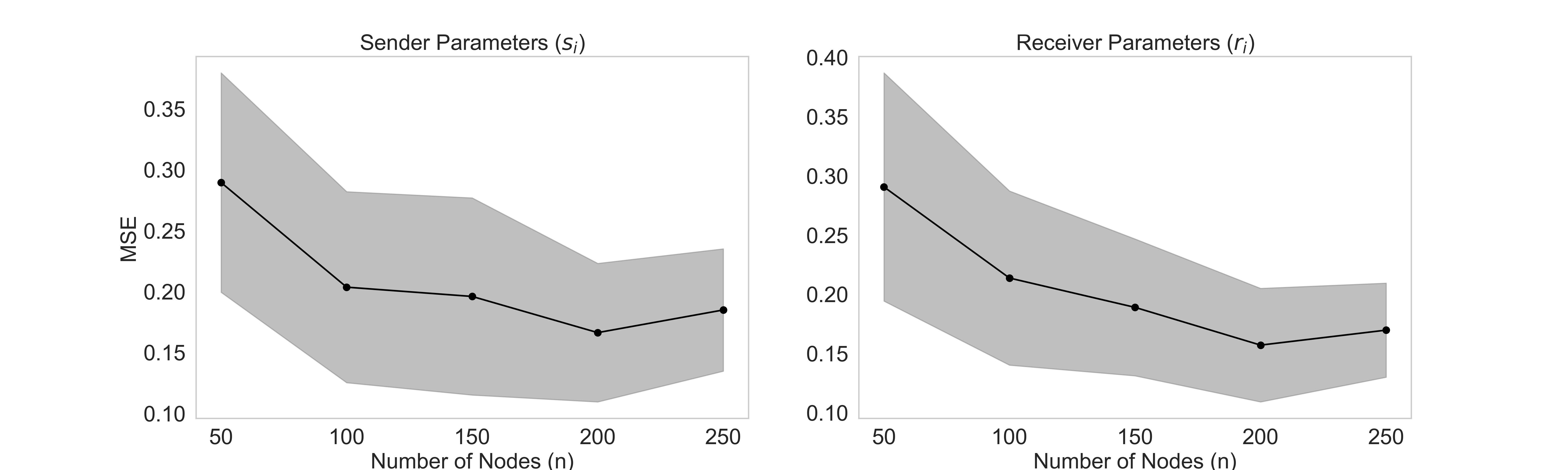} 
    \caption{The MSE of the sender parameters (left) and receiver parameters (right) for networks with $n \in \set{50, 100, 150, 200, 250}$. The points and the shaded regions represent the mean and one standard deviation over the 50 replicates, respectively.}
    \label{fig:sr-mse-increasing-nodes}
\end{figure}

\begin{figure}[htbp]
    \centering
    \includegraphics[width=\textwidth, keepaspectratio]{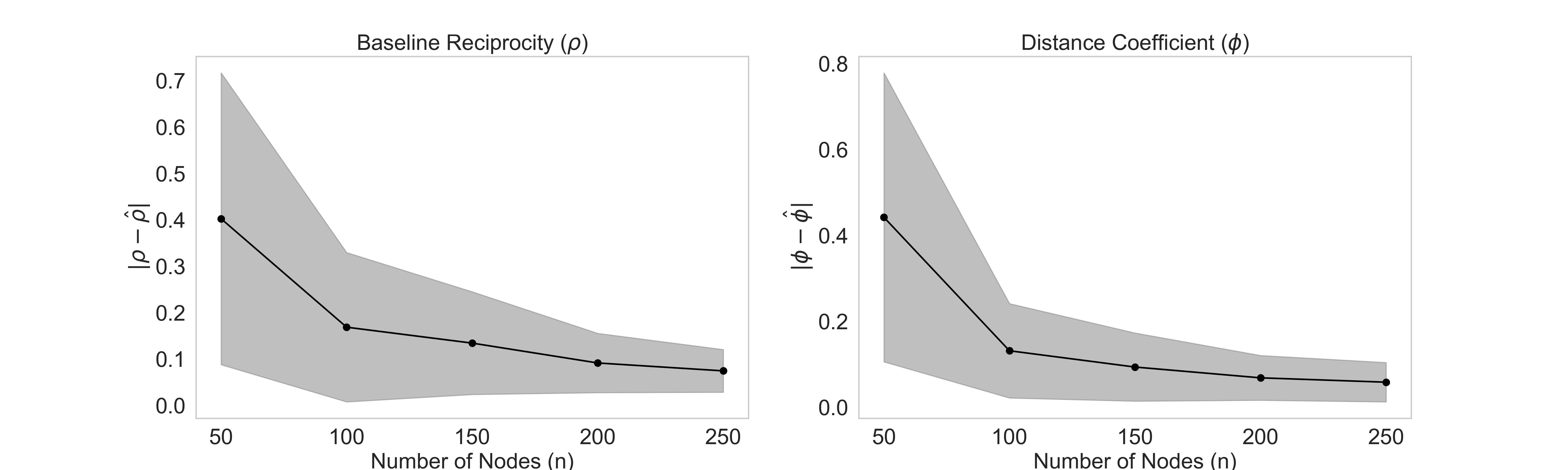} 
    \caption{Absolute deviations of the baseline reciprocity (left) and distance coefficient (right) for networks with $n \in \set{50, 100, 150, 200, 250}$. The points and the shaded regions represent the mean and one standard deviation over the 50 replicates, respectively.}
    \label{fig:rd-mse-increasing-nodes}
\end{figure}

\begin{figure}[htbp]
    \centering
    \includegraphics[height=0.3\textheight, width=\textwidth, keepaspectratio]{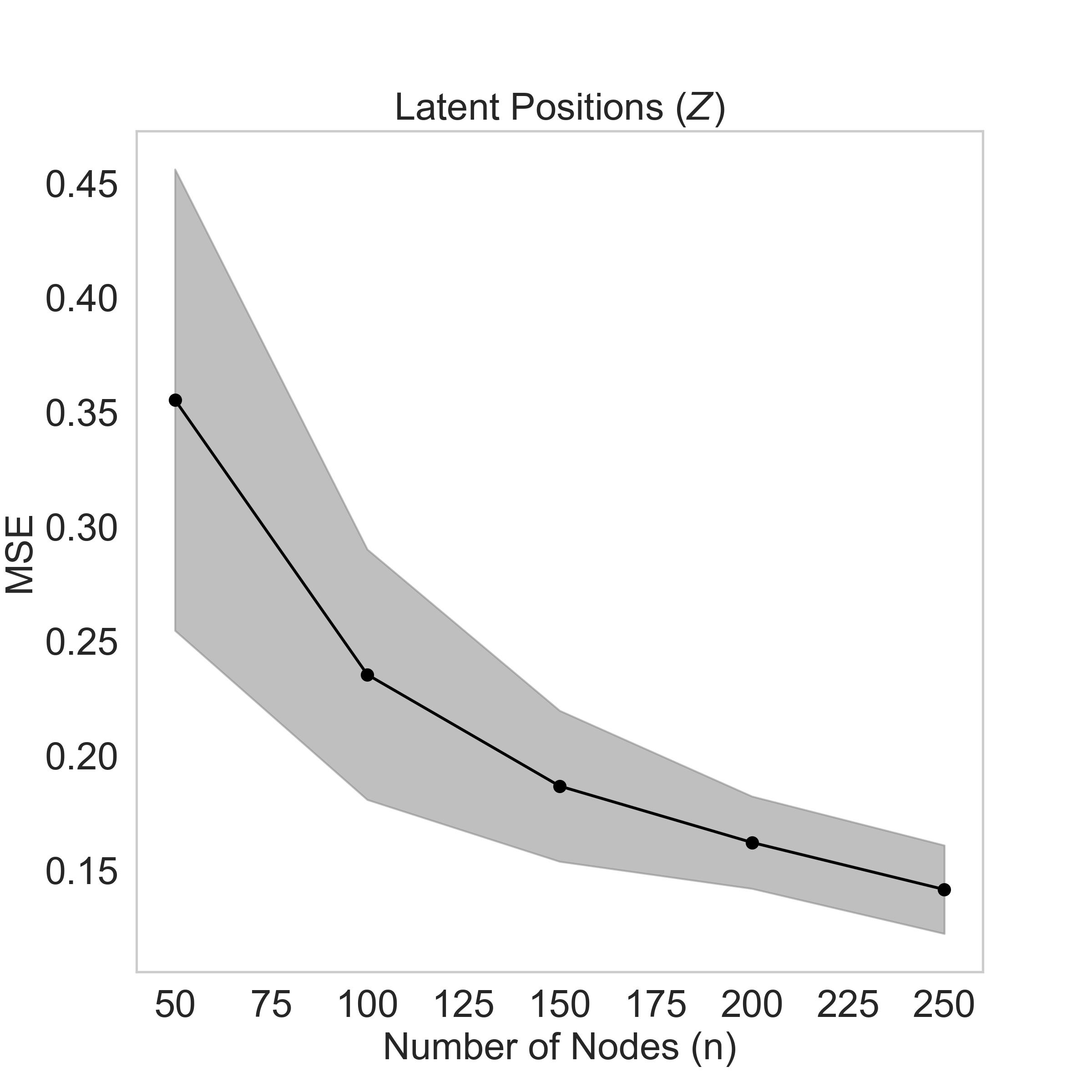} 
    \caption{MSE of the latent positions for networks with $n \in \set{50, 100, 150, 200, 250}$. The points and the shaded regions represent the mean and one standard deviation over the 50 replicates, respectively.}
    \label{fig:z-mse-increasing-nodes}
\end{figure}

We used the following metrics to assess the proposed estimators' accuracy. For the sender and receiver parameters, we used the following means squared errors (MSEs): $n^{-1} \sum_{i=1}^n (s_i - \hat{s}_i)^2$ and $n^{-1} \sum_{i=1}^n (r_i - \hat{r}_i)^2$. For the reciprocity parameters, we used the absolute deviations $\abs{\rho - \hat{\rho}}$ and $\abs{\phi - \hat{\phi}}$. For the latent positions, we calculated the MSE adjusted for the latent positions' invariance, i.e., we calculated $(nd)^{-1} \min_{\bO \in \mathcal{O}_d} \sum_{i=1}^n \norm{\bz_i - \bO \hat{\bz}_i}_2^2$, where $\mathcal{O}_d$ denotes the group of $d \times d$ orthogonal matrices.  Figures \ref{fig:sr-mse-increasing-nodes}--\ref{fig:z-mse-increasing-nodes} display the results for the sender and receiver parameters, reciprocity parameters, and the latent positions, respectively. The results indicate that as the number of nodes increases, the errors steadily decrease, indicating consistent recovery of the model parameters as $n$ grows.

\section{Real Data Applications}\label{sec:applications}

\subsection{The Network Data Sets}

To demonstrate how the proposed methodology can infer distance-dependent reciprocity, we consider three real-world social networks summarized in Table~\ref{tab:net_stats}. The first network involves advice relationships between lawyers at a law firm, the second network is an information-sharing network among employees at a manufacturing company, and the third network is a friendship network among high school students. These networks are all characterized by high levels of global reciprocity with each network having over 35\% of the dyadic relationships being mutual; however, we will use the proposed methodology to reveal that each network possesses a distinct heterogeneous reciprocity pattern.

\begin{table*}[ht]
\centering\small
    \begin{tabular*}{\textwidth}{@{\extracolsep\fill}lccc@{\extracolsep\fill}} \toprule
        & \textbf{Advice} & \textbf{Information-Sharing} & \textbf{Friendship} \\
\midrule
    Number of Nodes & 71 & 77 & 70 \\[0.25em]
    Density & 0.18 & 0.13 & 0.076 \\[0.25em]
    Percent of Mutual Ties & 39\% & 62\% & 50\% \\
\bottomrule
\end{tabular*}
\caption{Summary statistics for the three networks used in the real data applications.}
\label{tab:net_stats}
\end{table*}

To determine the type of reciprocity present in these different types of social interactions, we fit the three LSMs in Table~\ref{tab:model_types}, which are all nested within the proposed model, to each network: (1) the edge-independent model of \citet{krivitsky2009} (i.e., $\rho = \phi = 0$), (2) the homogeneous reciprocity model (i.e., $\phi = 0$), and (3) the general distance-dependent reciprocity model. For all models, we used $d = 2$ latent dimensions in order to visualize the latent space. All models used the priors and hyperparameters described in Section~\ref{sec:priors} and were estimated using the MCMC algorithm outlined in Section~\ref{sec:hmc}, which was run for 5,000 iterations after a burn-in of 2,500 iterations.

\subsection{Lazega Lawyers Advice Network}\label{subsec:lawyer}

Our first example is constructed from a data set of corporate law partnerships carried out in a Northeastern US law firm by \citet{lazega2001}. Three networks between $n = 71$ lawyers were constructed based on recorded strong-coworker, advice, or friendship relationships. In this work, we focus on the advice network, so $y_{ij} = 1$ indicates that lawyer $i$ went to lawyer $j$ for basic professional advice; otherwise, $y_{ij} = 0$. In addition to the network information, features of the individual lawyers are recorded, including their office location. This network has been previously analyzed using various edge-independent latent space models~\citep{ma2020, zhang2020}.

Our analysis aims to determine the type of reciprocity present in the advice network based on the three estimated network models. To do so, we calculated three commonly used information criteria: the Akaike information criterion (AIC), the Bayesian information criterion (BIC), and the deviance information criterion (DIC). The results are summarized in Table~\ref{tab:selection_criteria}, where lower values indicate a better fit to the observed network. For the advice network, all criteria selected the model with homogeneous reciprocity, which indicates that the local densities induced by the latent space under the edge-independent model are not sufficient to explain the amount of reciprocity in the network.

\begin{table}[t]
\centering\small
    \begin{tabular*}{\textwidth}{@{\extracolsep\fill}llcccc@{\extracolsep\fill}} \toprule
        \textbf{Network} & \textbf{Model} & \textbf{AIC} & \textbf{BIC} & \textbf{DIC} \\
\midrule
          & Edge-Independent~\citep{krivitsky2009} & 3013 & 4861 & 2931  \\[0.25em]
    Advice & Homogeneous &  {\bf 3009} & {\bf 4667} & {\bf 2925} \\[0.25em]
          & Distance-Dependent & 3015 & 4679 & 2927   \\
\midrule
          & Edge-Independent~\citep{krivitsky2009} & 2162 & 4218 & 1997  \\[0.25em]
    Information-Sharing   & Homogeneous &  2176 & 4024 & 1945 \\[0.25em]
          & Distance-Dependent & {\bf 2158} & {\bf 4012} & {\bf 1941}  \\
\midrule
          & Edge-Independent~\citep{krivitsky2009} & 1682 & 3497 & 1453  \\[0.25em]
    Friendship   & Homogeneous &  1685 & 3311 & 1464  \\[0.25em]
          & Distance-Dependent & {\bf 1641} & {\bf 3273} & {\bf 1421}  \\
\bottomrule
\end{tabular*}
\caption{The competing models' information criteria on the networks used in the applications. The value of the best performing model is bolded.}
\label{tab:selection_criteria}
\end{table}

\begin{figure}[tb]
\centering 
\includegraphics[height=0.2\textheight, width=\textwidth, keepaspectratio]{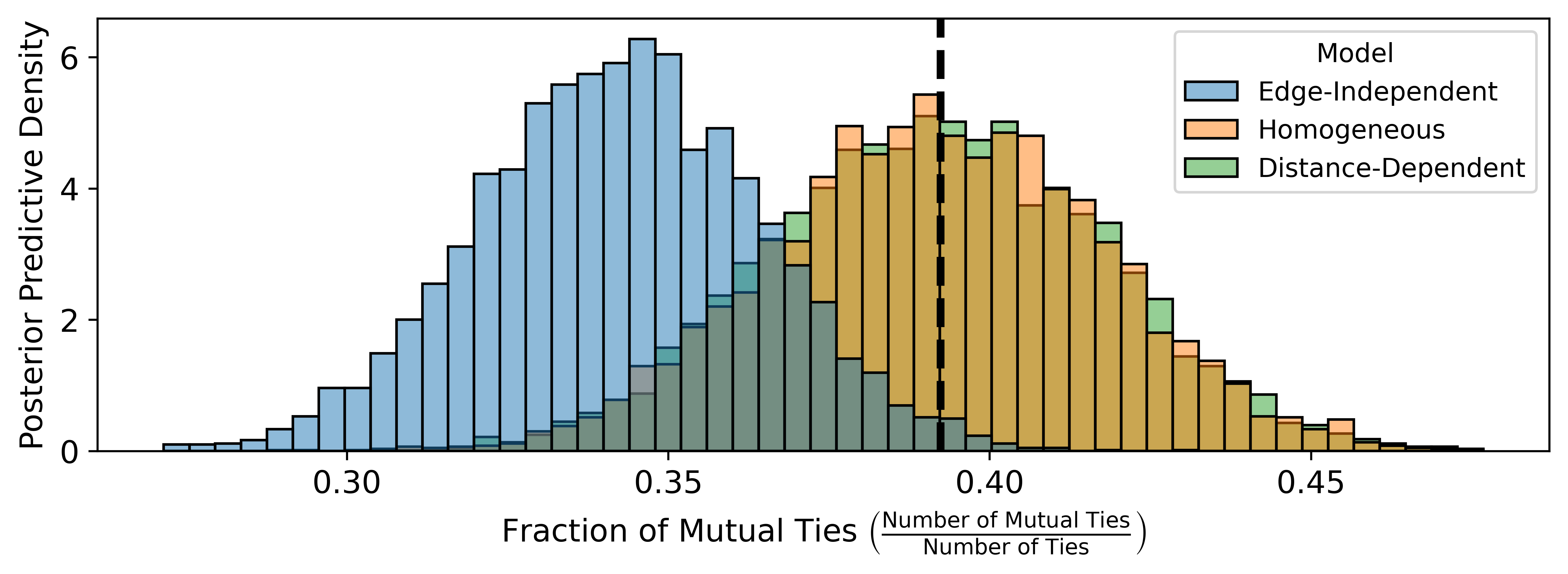}
    \caption{Posterior predictive densities of the fraction of mutual ties for the three competing models on the advice network. The dashed line denotes the fraction of mutual ties in the observed advice network.} 
\label{fig:advice_gof}
\end{figure}

The standard assumptions that justify information criterion for model selection are not met for LSMs, e.g., the number of parameters grows with $n$, so it is worth evaluating other goodness-of-fit criteria. In this context, a natural alternative is to compare the three models' ability to reconstruct the amount of reciprocity in the observed network. As such, we compared the three models' posterior predictive distributions, which is a popular approach for evaluating the goodness-of-fit of network models~\citep{hunter2008}. Figure~\ref{fig:advice_gof} displays the posterior predictive densities for the fraction of mutual ties, which demonstrate a clear preference for the proposed dyad-independent models. Specifically, the edge-independent model severely underestimates the fraction of mutual ties in the network, while the observed statistic falls well within the dyad-independent models' posterior predictive densities.

Next, we use the proposed model to determine the stochastic tendencies present in the advice network. Based on these goodness-of-fit measures, we used the homogeneous model for this inference. Figure~\ref{fig:lawyer_recip_ci} displays the marginal posterior distributions for the reciprocity parameters under the homogeneous model. The 95\% credible for $\rho$ is (0.43, 1.16), which indicates higher odds of observing a mutual tie than an unreciprocated tie even when controlling for the latent space. Figure~\ref{fig:lawyer_ls} shows the posterior mean estimates of the latent positions under the homogeneous reciprocity model. There is a clear separation of the lawyers from different offices, which indicates an increase in the propensity of advice sharing to occur between lawyers in the same office. This latter finding matches previous analyses using edge-independent LSMs~\citep{zhang2020}.

\begin{figure}[tb]
\centering
\includegraphics[height=0.2\textheight, width=\textwidth, keepaspectratio]{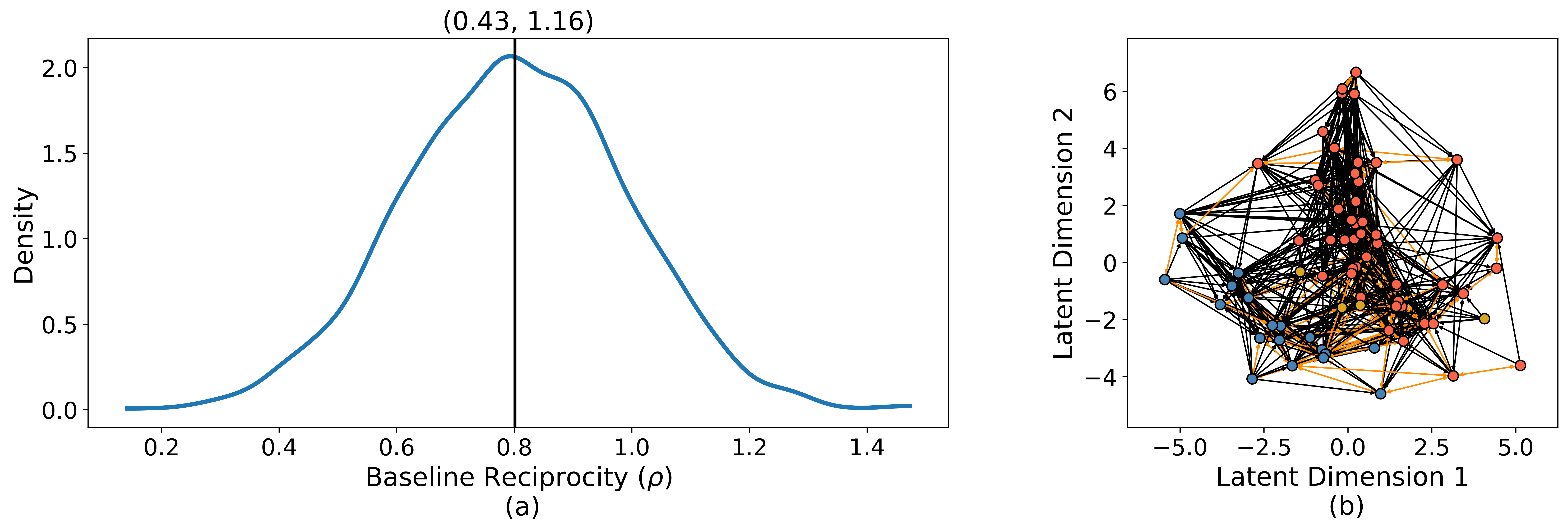}
    \caption{(a) The marginal posterior density of the reciprocity parameter $\rho$ under the homogeneous reciprocity model ($\phi = 0$) for the advice network. The 95\% credible interval is displayed on the top and the vertical line is the posterior mean. (b) The estimated latent positions under the homogeneous reciprocity model. The observed edges are displayed and the orange edges denote mutual ties (``$i \leftrightarrow j"$). Nodes are colored by office location: (Red) Boston, (Blue) Hartford, and (Orange) Providence.}
\label{fig:advice_cis}
\phantomlabel{a}{fig:lawyer_recip_ci}
\phantomlabel{b}{fig:lawyer_ls}
\end{figure}

\subsection{Intra-Organizational Information-Sharing Network}\label{sec:infoshare}

In our second application, we analyzed intra-organizational relationships at a European manufacturing company collected by \citet{cross2004}. The relational data was gathered by asking $n = 77$ employees on a research team to ``Please indicate the extent to which the people listed below provide you with information you use to accomplish your work'' according to the following scale: 0: I Do Not Know This Person/I Have Never Met This Person; 1: Very Infrequently; 2: Infrequently; 3: Somewhat Infrequently; 4: Somewhat Frequently; 5: Frequently; and 6: Very Frequently. We constructed a binary directed network from this data by recording $y_{ij} = 1$ if employee $i$ indicated that employee $j$ provided them with information at least somewhat frequently; otherwise, we set $y_{ij} = 0$. The data set also contains information on the employees, such as their office location.

\begin{figure}[t]
\centering 
\includegraphics[height=0.25\textheight, width=\textwidth, keepaspectratio]{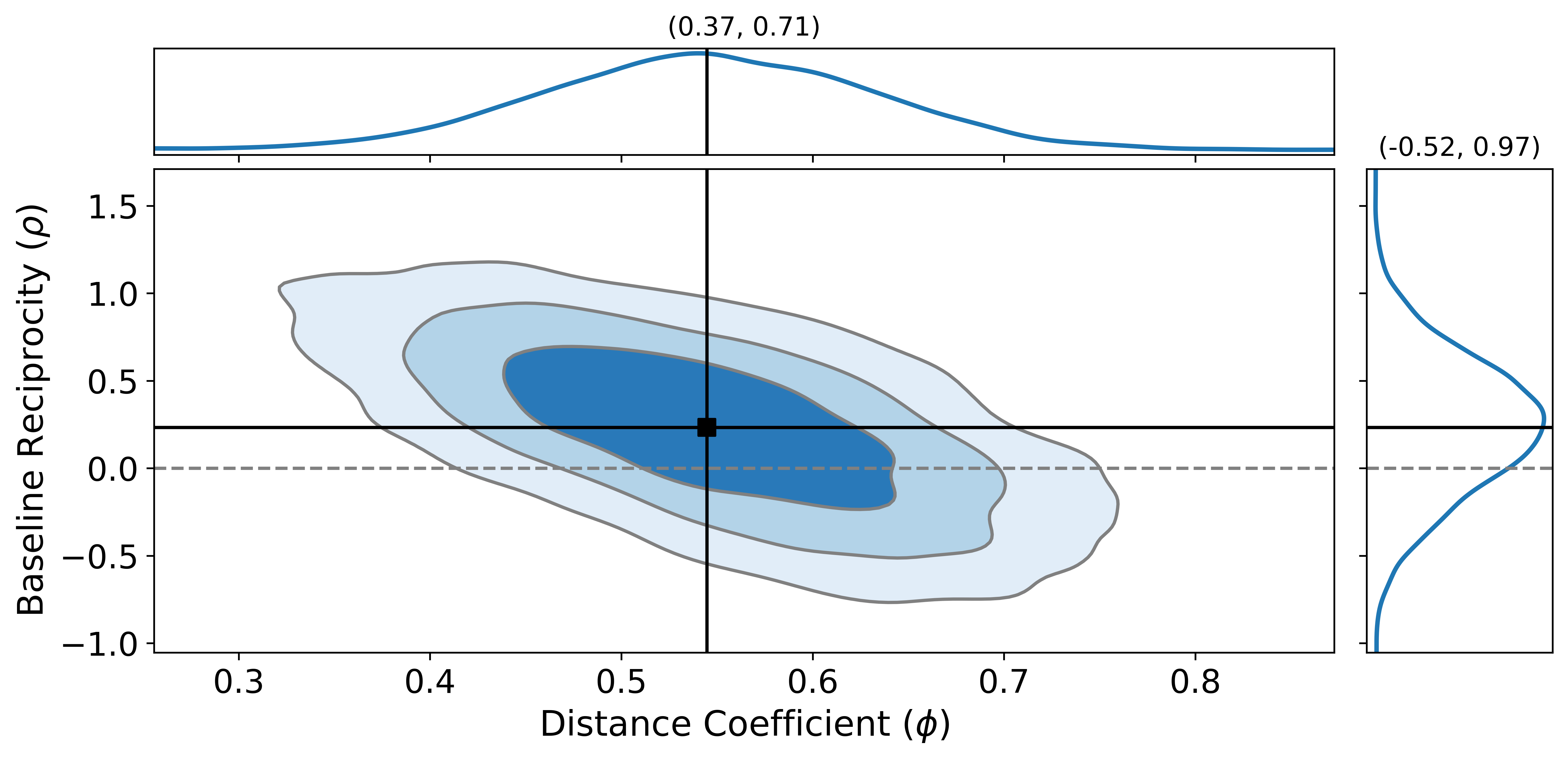}
    \caption{The joint posterior density of the reciprocity parameters $\rho$ and $\phi$ for the information-sharing network. The progressively lighter blue regions denote 50\%, 80\%, and 95\% credible regions, respectively. The marginal posterior densities and 95\% credible intervals are displayed on the top ($\phi$) and right ($\rho$) of the figure. The solid lines denote the posterior means and the dashed line correspond to $\rho$ equaling zero.}
\label{fig:consult_recip_cis}
\end{figure}

Our goal is to determine the reciprocity pattern present in the information-sharing network. All information criteria in Table~\ref{tab:selection_criteria} indicate that the model with distance-dependent reciprocity best fit the data. To determine the type of distance-dependent reciprocity, we used the joint posterior distribution of the baseline reciprocity $\rho$ and distance coefficient $\phi$ under the distance-dependent reciprocity model displayed in Figure~\ref{fig:consult_recip_cis}. The 95\% credible interval for $\phi$ is (0.37, 0.71) and its posterior mean $\hat{\phi} = 0.54$. Furthermore, the estimated posterior probability that $\phi$ is positive and within the unit interval, that is, $\hat{\mathbb{P}}(0 < \phi < 1 \mid \bY)$, equals one. As such, we conclude that the stochastic tendency towards mutual ties increases with the latent distance between two nodes. Moreover, this network demonstrates the phenomenon described in Section~\ref{sec:existing}, wherein reciprocation is likely even when edge formation is unlikely, which edge-independent LSMs cannot capture. To further interpret this result, we visualize the latent space.

Figure~\ref{fig:consult_ls} displays the inferred latent space from the distance-dependent model. The latent positions form four distinct clusters around the four office locations in Paris, Frankfurt, Warsaw, and Geneva. As such, the positive value of $\phi$ within the unit interval indicates that although information-sharing is less likely to occur between offices in the organization, see Equation~(\ref{eq:rlsm_logit}), the information-sharing that does occur between them is likely to be reciprocated, see Equation~(\ref{eq:dist_rep}). As described in Section~\ref{sec:existing}, existing edge-independent LSMs cannot capture this social phenomenon.

Lastly, we empirically verified that the inferred distance-dependent reciprocity pattern is present in the observed network. Figure~\ref{fig:consult_rep} displays the amount of reciprocity as a function of latent distance. Specifically, for each observed dyad $d_{ij} = (y_{ij}, y_{ji})$, we estimated the associated nodes' latent distance  $\hat{r}_{ij} = \norm{\hat{\bz}_i - \hat{\bz}_j}_2$, where the estimates correspond to the posterior means. We then estimated the local log odds ratio in Equation~(\ref{eq:dist_rep}) for dyads whose distance fell within a given window $\ell \leq \hat{r}_{ij} \leq u$, that is, we calculated $\hat{\rho}_{[\ell,u]} = \log\{(n_{i-j} n_{i\leftrightarrow j}) / (n_{i \leftarrow j} n_{i \rightarrow j})\}$, where the counts refer to the number of dyads of a given type whose latent distance lies within the window $\ell \leq \hat{r}_{ij} \leq u$. Figure~\ref{fig:consult_rep} plots the local log odds ratio estimates using a sliding window of width two with endpoints successively shifted by 0.5, that is, $\hat{\rho}_{[0,2]}, \hat{\rho}_{[0.5, 2.5]}, \hat{\rho}_{[1, 3]}$, and so on. We only reported estimates for windows that included at least one of each dyad type. It is clear from this plot that reciprocity does increase with latent distance.

\begin{figure}[t]
\centering 
\includegraphics[height=0.25\textheight, width=\textwidth, keepaspectratio]{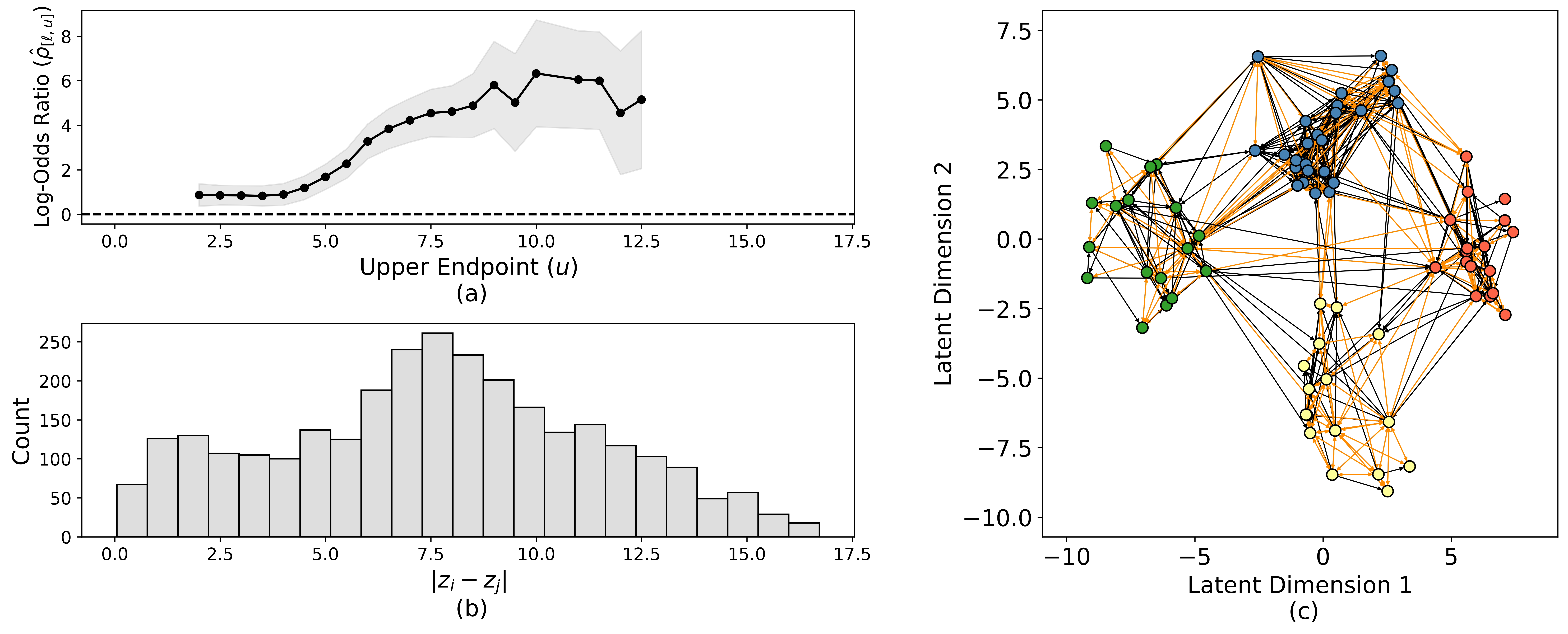}
    \caption{Latent space summaries for the information sharing network. (a) The black curve indicates the estimated local log odds ratio $\hat{\rho}_{[\ell, u]}$ and the shaded regions are 95\% point-wise confidence intervals. (b) Histogram of estimated latent distances under the distant-dependent reciprocity model. (c) The estimated latent positions under the distant-dependent reciprocity model. The observed edges are displayed and the orange edges denote mutual ties (``$i \leftrightarrow j"$). Nodes are colored by office location: (Red) Paris, (Blue) Frankfurt, (Yellow) Warsaw, (Green) Geneva.}
    \phantomlabel{a}{fig:consult_rep}
    \phantomlabel{b}{fig:consult_dist}
    \phantomlabel{c}{fig:consult_ls}
\end{figure}

\subsection{Illinois High School Friendship Network}

Lastly, we return to the friendship network used to motivate distance-dependent reciprocity in Section~\ref{sec:intro}. The network was constructed from observed friendship relations among $n = 70$ male students at a high school in Illinois by \citet{coleman1964}. The students were asked, ``What fellows here in school do you go around with most often?'' in the Fall and Spring semesters of the 1957 to 1958 school year. We formed a directed network by recording $y_{ij} = 1$ if student $i$ named student $j$ as a friend in either semester and $y_{ij} = 0$ otherwise.

\begin{figure}[t]
\centering 
\includegraphics[height=0.23\textheight, width=\textwidth, keepaspectratio]{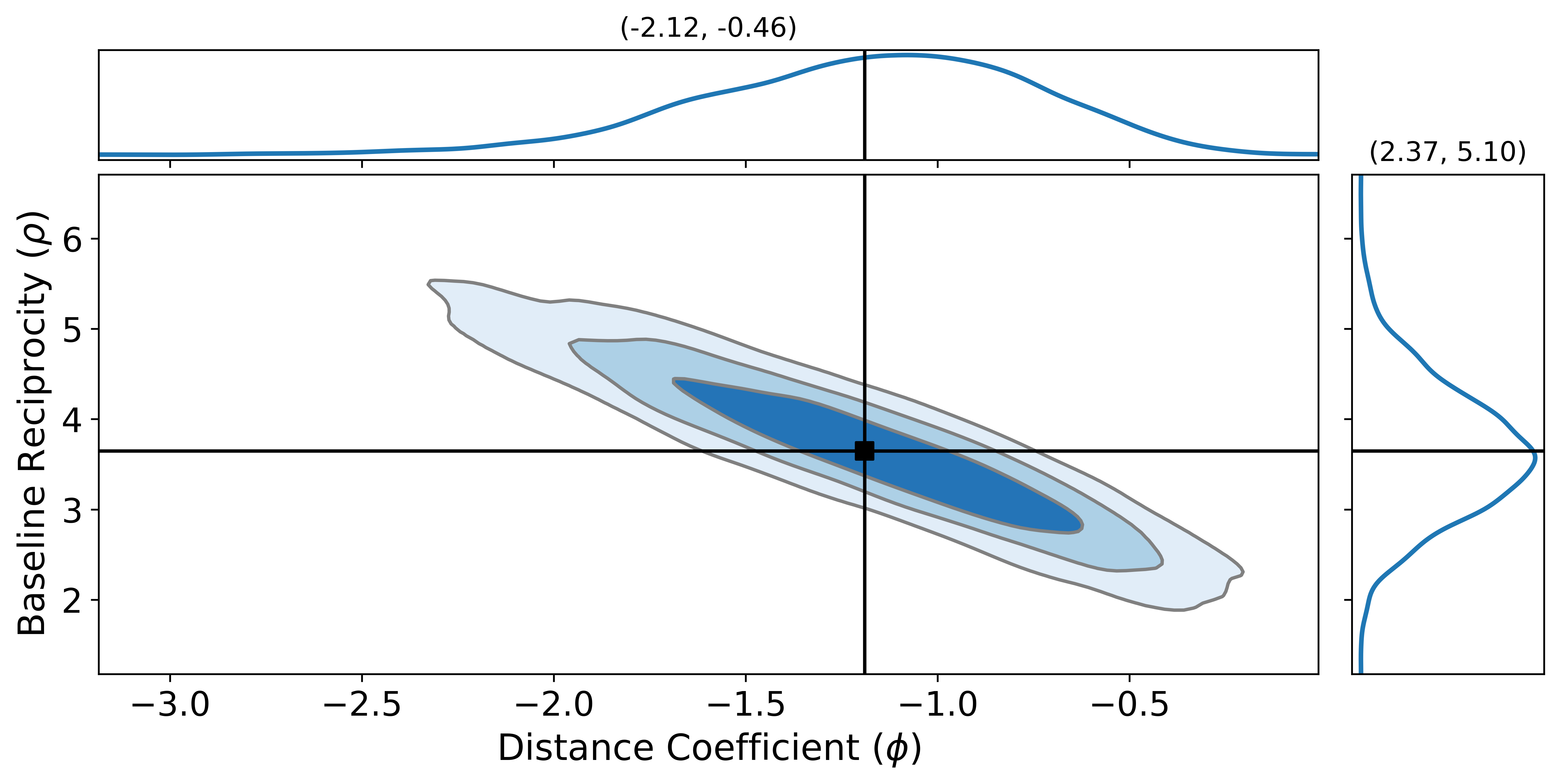}
\caption{The joint posterior density of the reciprocity parameters $\rho$ and $\phi$ for the high school friendship network. The progressively lighter blue regions denote 50\%, 80\%, and 95\% credible regions, respectively. The marginal posterior densities and 95\% credible intervals are displayed on the top ($\phi$) and right ($\rho$) of the figure. The solid lines denote the posterior means.}
\label{fig:friendship_recip_ci}
\end{figure}

All information criteria in Table~\ref{tab:selection_criteria} indicate that the distance-dependent reciprocity model best fit the friendship network. To determine the type of distance-dependent reciprocity, we interpreted the joint posterior distribution of the baseline reciprocity $\rho$ and distance coefficient $\phi$ under the distance-dependent reciprocity model displayed in Figure~\ref{fig:friendship_recip_ci}. The 95\% credible interval for $\phi$ is $(-2.12, -0.46)$, and its posterior mean $\hat{\phi} = -1.19$. Furthermore, the estimated posterior probability that $\phi$ is negative, that is, $\hat{\mathbb{P}}(\phi < 0 \mid \bY)$, equals one. As such, we conclude that the stochastic tendency towards mutual ties decreases with the latent distance between two nodes. In other words, friendship relations are more likely to occur and be reciprocated between students with similar latent positions. This is the opposite of the effect in the information-sharing network.

\begin{figure}[t]
\centering 
\includegraphics[height=0.25\textheight, width=\textwidth, keepaspectratio]{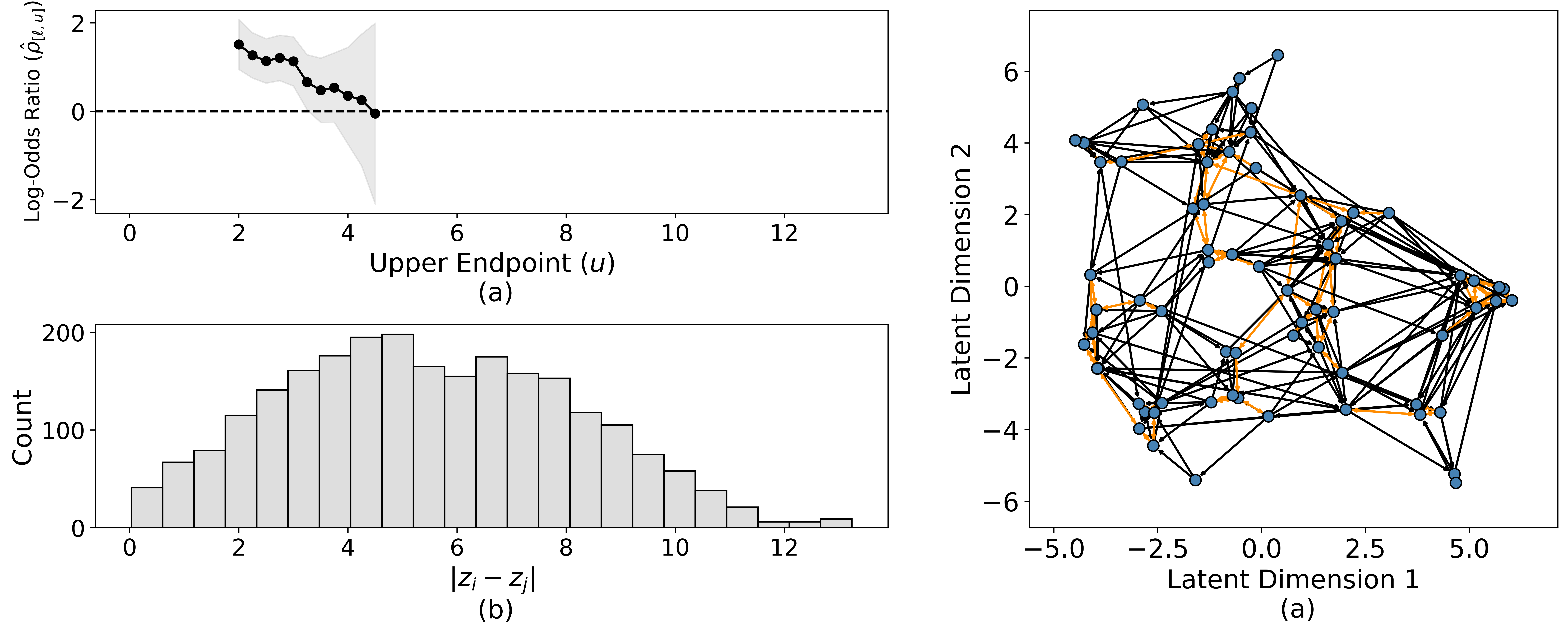}
    \caption{Latent space summaries for the high school friendship network. (a) The black curve indicates the estimated local log odds ratio $\hat{\rho}_{[\ell, u]}$ and the shaded regions are 95\% point-wise confidence intervals. (b) Histogram of estimated latent distances under the distant-dependent reciprocity model. (c) The estimated latent positions under the distant-dependent reciprocity model. The observed edges are displayed and the orange edges denote mutual ties (``$i \leftrightarrow j"$).}
    \phantomlabel{a}{fig:friend_rep}
    \phantomlabel{b}{fig:friend_dist}
    \phantomlabel{c}{fig:friend_ls}
\end{figure}

To further interpret this result, we visualized the inferred latent space under the distance-dependent reciprocity model in Figure~\ref{fig:friend_ls}. This plot shows mutual connections occur more frequently over short distances in latent space. As described in the previous section, we empirically checked this observation by plotting the local log odds ratio as a function of the latent distances in Figure~\ref{fig:friend_rep}. In this case, we used a sliding window of width two with endpoints successively shifted by 0.25. Note that the local log odds ratio is only calculated for windows that include all four dyad types, which are unavailable once $\hat{r}_{ij}$ is roughly greater than 5. This plot confirms that the local log odds ratio decreases with latent distance. In conclusion, these results indicate that the presence of friendships and their reciprocation are both increased if the two students have similar latent positions.

\section{Discussion}\label{sec:discussion}

In this article, we presented a novel latent space model for inferring distance-dependent reciprocity in directed networks, which is a characteristic of relationships that previous statistical network models could not infer. As such, we were able to use the proposed methodology to uncover the presence of different distance-dependent reciprocity types in three real-world networks from the social sciences. There are various ways to improve the proposed methodology. First, the proposed MCMC algorithm does not scale to large networks, so exploring the performance of faster inference schemes, such as variational inference~\citep{wainwright2008}, would be beneficial. Second, the methodology only applies to unweighted networks; however, one may be interested in modeling the correlation between edge pairs in weighted network data. Using a bivariate normal observation model is a possible solution. Lastly, the odds ratio in Equation~(\ref{eq:dist_rep}) may depend on the latent distances in a non-linear manner, which could be modeled using a nonparametric approach, such as a spline approximation. We leave these extensions to future work.

\singlespacing
\bibliographystyle{apalike}
\bibliography{reference}

\end{document}